\documentclass[aps, showpacs, amsmath,twocolumn,amssymb,floatfix,nofootinbib,prb]{revtex4}
\usepackage{graphicx} 
\usepackage{color}
\definecolor{Blue}{rgb}{0.3,0.3,0.9}
\usepackage[T1]{fontenc}
\usepackage{ae,aecompl}
\usepackage{pstricks}
\usepackage{pstricks-add}
\usepackage{epsfig}
\usepackage{epstopdf}

\newcommand{\vect}[1]{\boldsymbol{#1}}

\begin{document}


\title{Spin-orbit interactions in a helical Luttinger liquid with a Kondo impurity}


\author{Erik Eriksson}
\affiliation{Department of Physics, University of Gothenburg, SE-412 96 Gothenburg, Sweden}


\begin{abstract}
The combined effect of Rashba and Dresselhaus spin-orbit interactions on the physics of a helical Luttinger liquid coupled to a Kondo impurity is studied. A Rashba coupling can potentially destroy the Kondo singlet formation in certain parameter regimes [Phys.~Rev.~B 86, 161103(R) (2012)]. This effect is here shown to vanish for sufficiently strong Dresselhaus interaction. The transport properties of the system are investigated by calculating electrical conductance, current and current fluctuations, and thermal conductance.

\end{abstract}



\maketitle


\section{Introduction}

Topological insulators constitute a new phase of matter, with preserved time-reversal invariance and a bulk energy gap that is accompanied by topologically protected gapless states on the boundary.\cite{KM,bhz,Konig,Review1,Review2} These boundary states are helical, forming an odd number of Kramers pairs where electron spin is coupled to the direction of propagation. On the edge of a two-dimensional (quantum spin Hall) topological insulator,\cite{bhz,Konig} or in a topological insulator nanowire,\cite{ezy,peng,kong} such helical states will in the presence of electron-electron interactions give rise to the new class of gapless one-dimensional correlated electron liquid known as the {\it helical Luttinger liquid}.\cite{Wu,Xu}

This one-dimensional electron system is distinct from standard spinless, spinful and chiral Luttinger liquids,\cite{G,gnt} and from the "spiral Luttinger liquids" formed by a spin-orbit coupled quantum wire\cite{gritsev} in a magnetic field,\cite{alicea,bbs} due to its time-reversal invariance with a single Kramers pair. This property makes the helical liquid stable against weak disorder since it forbids single-electron backscattering from spinless impurities,\cite{KM} and although electron-electron interaction may cause impurity-induced correlated two-electron backscattering, the gapless helical liquid is still left intact provided the screening is not very weak.\cite{Wu,Xu} However, time-reversal invariance does not protect the helical liquid against the effects from magnetic impurities, occurring due to dopant ions or trapped bulk electrons caused by potential inhomogeneities, allowing electrons to backscatter via spin exchange. 

Magnetic impurities in conventional Luttinger liquids are well-studied.\cite{furusaki} Single-electron backscattering from non-magnetic impurities is a relevant perturbation causing perfect reflection for repulsive electron-electron interaction.\cite{KF1} The Kondo scattering off magnetic impurities leads to a low-energy behavior governed by a strong-coupling fixed point where the impurity is completely screened by the electrons, approached with the anomalous exponents of a non-Fermi liquid\cite{LT,FN,FJ} and resulting in vanishing zero-temperature conductance. For a helical Luttinger liquid, studied early on as a truncated model for a standard Luttinger liquid, there is only backward Kondo scattering and the Kondo effect was shown to map exactly onto that in a Fermi liquid.\cite{si} This yields the typical scaling behavior for the Kondo effect of ordinary conduction electrons.\cite{Hewson} Hence, for the helical Luttinger liquid with antiferromagnetic Kondo coupling there is a low-temperature strong-coupling fixed point, where the local magnetic moment (stabilized against charge fluctuations by the electron correlations\cite{ps}) is screened by the electrons effectively leaving behind a spinless impurity.\cite{Wu,M,M2} The edge of a two-dimensional topological insulator merely bypasses this Kondo singlet, thus restoring unitary conductance $G = G_0 = e^2 /h$ at zero temperature.\cite{M} Transport properties in the vicinity of this strong-coupling fixed point then follow power-laws governed by the correlated two-electron backscattering, whereas at high temperatures the Kondo scattering results in a logarithmic temperature correction in the linear conductance,\cite{M} which however vanishes in the dc limit.\cite{T}

In addition to the atomic spin-orbit coupling that lies behind the very existence of the quantum spin Hall state,\cite{bhz} with the idealized helical edge states with right (left) movers with spin up (down), there will also be additional spin-orbit couplings which make spin no longer conserved.\cite{brune} As long as the bulk gap remains, preserved time-reversal invariance ensures the robustness of the helical liquid.\cite{Wu,KM} However, an important question is how the Kondo physics of magnetic impurities is affected by the additional spin-orbit interactions in the helical liquid. For a quantum well with structure inversion asymmetry (SIA), i.e. an asymmetry in the confining potential along the direction perpendicular to the well, there will be a {\it Rashba} spin-orbit interaction.\cite{rashba,WinklerBook,erh,grundler,rothe} This interaction can be controlled by an external gate voltage, a scenario explored in Ref.~\onlinecite{essj}. When the crystal structure lacks inversion symmetry, i.e.~in the case of bulk inversion asymmetry (BIA), there will also be the {\it Dresselhaus} spin-orbit interaction.\cite{dresselhaus,WinklerBook,erh} Including this type of interaction hence adds a crucial piece to a fuller understanding of the effects of spin-orbit interactions in these systems. In fact, both HgTe and InAs have the inversion asymmetric zincblende crystal structure, thereby displaying the Dresselhaus interaction. HgTe/CdTe quantum wells, the first quantum spin Hall insulator experimentally observed,\cite{Konig} is known for its large Rashba coupling\cite{Buhmann} and one here expects the Dresselhaus coupling to be much smaller in comparison.\cite{silsbee} For InAs/GaSb quantum wells, where helical edge states also have been demonstrated,\cite{lhqwz,kds} the Dresselhaus interaction is sizeable\cite{cw,my} and realistic predictions for spin-orbit effects must take it into account.

It is therefore the aim of this paper to revisit the problem of a Kondo impurity in a helical Luttinger liquid with spin-orbit interaction. The results announced in Ref.~\onlinecite{essj} are extended to include both Rashba and Dresselhaus interactions, showing that the obstruction of Kondo singlet formation in certain parameter regimes persists but will vanish for sufficiently strong Dresselhaus interaction. The linear conductance far above the Kondo temperature is obtained, also in the low-frequency limit where a rate-equation approach must be employed.\cite{T} Furthermore, current-voltage and noise characteristics of the backscattered current are calculated, as well as the expression for the thermal conductance, showing how the electric-field adjustable Rashba interaction offers a mechanism to control transport properties.

\section{Model}

The low-energy dynamics of the helical electrons, neglecting interactions, is described by the one-dimensional Dirac Hamiltonian
\begin{equation} \label{H0}
H_0 = v_F \int \textrm{d}x \ \Psi^{\dagger}(x) \left[ -i \sigma^{z} \partial_x \right] \Psi (x) 
\end{equation}
where $v_F$ is the Fermi velocity, and the two-component spinor $\Psi = \big( \psi_{\uparrow}, \psi_{\downarrow} \big)^T$. Here $ \psi_{\uparrow} \, (\psi_{\downarrow})$ annihilates an electron with spin up (down) in the $z$ direction, the growth direction of the quantum well. Unless stated otherwise, the units are such that $\hbar = k_B =1$ and the lattice constant $a =1$.

Rashba and Dresselhaus spin-orbit interactions, of strength $\alpha$ and $\beta$ respectively, to leading order add the terms\cite{dresselhaus,rashba,WinklerBook,rothe}
\begin{eqnarray}
H_R &=&  \alpha \int \textrm{d}x \ \Psi^{\dagger}(x) \left[ -i\sigma^{y} \partial_x \right] \Psi (x) \label{R} \\
H_D &=& \beta  \int \textrm{d}x \ \Psi^{\dagger}(x) \left[ -i\sigma^{x} \partial_x \right] \Psi (x)  \label{D}
\end{eqnarray}
to the Hamiltonian. The Kondo interaction, describing the antiferromagnetic exchange coupling between the electrons and a spin-1/2 magnetic impurity located at $x=0$, is given by
\begin{equation} \label{HK}
H_K =  \Psi^{\dagger} (0) \left[  J_x \sigma^x S^x + J_y \sigma^y S^y + J_z \sigma^z S^z \right] \Psi(0),
\end{equation}
where $\sigma^{i}$ ($S^{i}$), $ i=x,y,z$, are the electron Pauli matrices (impurity spin operators). At a quantum well interface, spin-orbit induced magnetic anisotropy for an impurity would result in\cite{Ujsaghy,Zitko} $J_x = J_y \neq J_z$.

The effect of a Rashba interaction (\ref{R}) on the helical states (\ref{H0}) was studied in Ref.~\onlinecite{VO}, and that analysis can now be adapted to also include the Dresselhaus coupling (\ref{D}). Since $H_0 + H_R + H_D$ is given by
\begin{equation}
 \int \textrm{d}x \ \Psi^{\dagger}(x) \left[ -i \left( v_F\sigma^{z} + \alpha \sigma^{y} + \beta \sigma^{x} \right) \partial_x \right] \Psi (x) ,
\end{equation}
it follows that a rotation $\Psi \rightarrow \Psi' =  e^{i\sigma^y \phi/2}  e^{-i\sigma^x \theta/2} \Psi$ diagonalizes $H_0 + H_R + H_D \rightarrow H'_0$, with
\begin{equation} \label{H0p}
H'_0 = v_{\alpha \beta} \int \textrm{d}x \ \Psi'^{\dagger}(x) \left[ -i \sigma^{z} \partial_x \right] \Psi' (x) , 
\end{equation} 
where $v_{\alpha \beta} = \sqrt{v_F^2 + \alpha^2 + \beta^2}$, and the rotation angles are determined by
\begin{eqnarray}
\sin \theta &=& \alpha / \sqrt{v_F^2 + \alpha^2}, \\
\sin \phi &=&  \beta / \sqrt{v_F^2 + \alpha^2 + \beta^2}.
\end{eqnarray}
Thus the spectrum of $H_0 + H_R + H_D$ is linear, and the components $ \psi'_{+}$ and $ \psi'_{-}$ of the spinor $\Psi'$ are left- and right movers. It is therefore possible to treat the effects of the spin-orbit interactions (\ref{R})-(\ref{D}) exactly by working in the rotated basis. 

The Kondo interaction (\ref{HK}) is now analyzed in the rotated basis. In order to have the impurity and electron spins quantized along the same axis, rotate $\vect{S} \rightarrow \vect{S}^{\prime} =e^{iS^y \phi/2} e^{-i S^x \theta/2}\vect{S} e^{iS^x \theta/2}e^{-iS^y \phi/2}$. Then the Kondo Hamiltonian becomes
\begin{eqnarray} 
H'_K &=& \Psi'^{\dagger} (0) \ [\ J'_x \sigma^{x} S'^{x} + J'_y \sigma^{y} S'^{y} + J'_z \sigma^{z} S'^{z} \nonumber \\
&&   + J_{xy}  \sigma^{x} S'^{y} +  J_{yx} \sigma^{y} S'^{x}     
+  J_{xz} \sigma^{x} S'^{z}   \label{HKp} \\   
&& +  J_{zx} \sigma^{z} S'^{x}    +  J_{yz}  \sigma^{y} S'^{z} +  J_{zy} \sigma^{z} S'^{y}      \   ]\  \Psi'(0) ,  \nonumber  
\end{eqnarray}
where the coupling constants are given by
\begin{align}
& J'_x = J_x \cos^2 \phi + J_y \sin^2 \theta \sin^2 \phi + J_z \cos^2 \theta \sin^2 \phi, \nonumber \\
& J'_y = J_y \cos^2 \theta + J_z \sin^2 \theta , \nonumber \\
& J'_z = J_x \sin^2 \phi + J_y \sin^2 \theta \cos^2 \phi + J_z \cos^2 \theta \cos^2 \phi, \nonumber \\
& J_{xy} =  J_{yx} = (J_z - J_y) \cos \theta \sin \theta \sin \phi, \nonumber \\
& J_{xz} = J_{zx} = (J_x - J_y \sin^2 \theta - J_z \cos^2 \theta) \cos \phi \sin \phi, \nonumber \\
& J_{yz} = J_{zy} = (J_y - J_z) \cos \theta \sin \theta \cos \phi. \nonumber 
\end{align}

The next step is to include the electron-electron interactions allowed by time-reversal symmetry. Assuming a band away from half-filling, $k_F \neq \pi/2$, Umklapp scattering
\begin{eqnarray} \label{umkl}
H'_{um} &=& g_{um} \int \textrm{d}x \ e^{-i 4 k_F x}  \psi_{+}'^{\dagger}(x)  \psi_{+}'^{\dagger}(x+a) \nonumber \\
&& \qquad \quad \times \psi'_{-}(x)  \psi'_{-}(x+a) + H.c. 
\end{eqnarray} 
can be ignored.\cite{Wu} Here a point splitting with the lattice constant has been performed. Dispersive and forward scattering are given by
\begin{eqnarray} \label{d}
H'_{d} &=& g_{d} \int \textrm{d}x \  \psi_{+}'^{\dagger}(x)  \psi_{+}'(x) \psi'^{\dagger}_{-}(x)  \psi'_{-}(x) , \\
H'_{f} &=& \frac{g_{f}}{2} \sum_{s = \pm} \int \textrm{d}x \  \psi_{s}'^{\dagger}(x)  \psi_{s}'(x) \psi'^{\dagger}_{s}(x+a)  \psi'_{s}(x+a) , \nonumber \\ \label{f}
\end{eqnarray} 
and at the impurity site there will in addition be correlated two-particle backscattering \cite{Wu,Xu,crepin}
\begin{equation} \label{bs}
H'_{2p} = g_{2p}  \psi_{+}'^{\dagger}(0)  \psi_{+}'^{\dagger}(a) \psi'_{-}(0)  \psi'_{-}(a) + H.c. 
\end{equation} 
and inelastic single-particle backscattering \cite{Schmidt,Lezmy,Budich}
\begin{eqnarray} 
H'_{ie} &=& g_{ie}  \left( \psi_{+}'^{\dagger}(0)  \psi_{+}'(0) -  \psi_{-}'^{\dagger}(0)  \psi_{-}'(0) \right) \nonumber \\
&&  \qquad \quad \times \psi_{+}'^{\dagger}(a)   \psi'_{-}(a) + H.c.  \label{ie}
\end{eqnarray}

Employing standard bosonization methods \cite{G,gnt} the interacting Hamiltonian $H'_0 + H'_d + H'_f$ forms a helical Luttinger liquid
\begin{equation} \label{HLL}
H'_{HLL} = \frac{v}{2} \int \textrm{d}x \left[ (\partial_x\varphi)^2 +
(\partial_x \vartheta)^2\right] ,
\end{equation}
where the electron operators $\psi_{\pm} = (2\pi \kappa)^{-1/2} e^{-i \sqrt{\pi}(\vartheta \pm \varphi)}$ are represented in terms of non-chiral Bose fields $\varphi (x) $ and $\vartheta (x) $, with $[\varphi (x) , \partial_y \vartheta (y)] = i \pi \delta (x-y) $. Here $\kappa \, \approx v_F/D$ is the edge state penetration depth acting as short-distance cutoff with $D$ the bandwidth. The velocity $v \!=\![(v_{\alpha\beta} \!+ \!g_f/\pi)^2\!-\!(g_d/\pi)^2]^{1/2}$, and the Luttinger parameter $K =  [(\pi v_{\alpha} \!+ \!g_f - g_d)/(\pi v_{\alpha} \!+ \!g_f + g_d)]^{1/2}$. The correlated two-particle and inelastic single-particle backscattering terms in Eqs.~(\ref{bs})-(\ref{ie}) are now expressed as
\begin{eqnarray}
H_{2p}^{\prime} \!&\!=\!&\! \frac{g_{2p}}{2(\pi \kappa)^2} \cos[\sqrt{16\pi K}\varphi(0)] \label{BackBoson} \\
H_{ie}^{\prime} \!&\!=\!&\! \frac{g_{ie}}{2\pi^2\sqrt{K} }:  \left[ \partial_x^2 \vartheta(0) \right]  \cos [ \sqrt{4\pi K}\varphi(0)   ] :      \label{IEBoson}
\end{eqnarray}
where $:...:$ denotes normal ordering. The Kondo interaction (\ref{HKp}) is represented as
\begin{eqnarray}
H_K^{\prime}\! \!&\!=\!& \!\!\frac{A}{\pi \kappa}\!\cos[\sqrt{4\pi K}\varphi (0)  ]\! +\! \frac{B}{\pi \kappa} \!\sin[\sqrt{4\pi K}\varphi (0)  ]  \nonumber \\
&& \qquad \qquad \qquad + 
\frac{C}{\sqrt{\pi K}}\partial_x\vartheta(0)  \label{KondoBoson}
\end{eqnarray}
where the operators $A$, $B$ and $C$ are defined as
\begin{eqnarray}
A &\equiv &  J'_x S'^x + J_{xy} S'^y + J_{xz} S'^z , \nonumber   \\
B &\equiv & J_{yx} S'^x + J'_y S'^y + J_{yz} S'^z  ,\label{ABC}\\
C &\equiv & J_{zx} S'^x + J_{zy} S'^y + J'_z S'^z.  \nonumber
\end{eqnarray}

\section{Kondo temperature}

The low-temperature behavior of the model can now be determined through a perturbative renormalization-group analysis. First, from Eqs.~(\ref{BackBoson}) and (\ref{IEBoson}) it is seen that the correlated two-particle and inelastic single-particle backscattering operators have scaling dimensions $4K$ and $K+2$, respectively. Hence, for Luttinger parameter $K < 2/3$, the two-particle backscattering is the dominating of the two. For $K< 1/4$ it turns relevant, with a crossover from weak to strong coupling at a temperature \cite{KaneFisher,M} $T_{2p} \approx Dg_{2p}^{1/(1-4K)}$.

Let us now derive the renormalization-group equations for the Kondo couplings in Eq.~(\ref{HKp}). The local partition function at the impurity site corresponding to the Hamiltonian $H'_{HLL} + H'_K$ is written as a functional integral
\begin{eqnarray}
\mathcal{Z} &=& \int \mathcal{D}[ \varphi]  \ e^{-S[\varphi]} \nonumber \\
&=&  \int \mathcal{D}[ \varphi]  \ \exp \left\{ - \frac{1}{2\pi} \int \mathrm{d} \omega \ |\omega | |\varphi (\omega) |^2 - S_K[ \varphi(\tau)]  \right\},  \nonumber
\end{eqnarray}
with the imaginary time action $S_K$ for the Kondo terms given by
\begin{eqnarray}
S_K [\varphi (\tau) ]  &=& \int_0^{\beta} \mathrm{d} \tau \ \Big\{ \frac{A}{\pi \kappa}\!\cos[\sqrt{4\pi K}\varphi (\tau)  ]   \nonumber \\
&& \   + \frac{B}{\pi \kappa} \!\sin[\sqrt{4\pi K}\varphi (\tau)  ] +
\frac{i C\sqrt{K} }{\sqrt{\pi  v}}\partial_{\tau} \varphi(\tau) \Big\} \nonumber \\
\end{eqnarray}
and the impurity operators $A$, $B$ and $C$ given by Eq.~(\ref{ABC}). The analysis will be perturbative in the Kondo couplings, with $S_0 [\varphi] = S[\varphi] - S_K [\varphi]$ the unperturbed action. Although the $J'_z$ coupling may be treated exactly in the case of zero spin-orbit couplings,\cite{M2} the additional coupling terms introduced in the Hamiltonian (\ref{HKp}) by the spinor rotation precludes this line of attack. Since the exact analysis in $J'_z$ yields different physics only for $J'_z /(\pi v) > 2(K+\sqrt{K})$, where the Kondo effect disappears,\cite{M2} for reasonable strength of the electron-electron interaction the perturbative treatment of $J'_z$ is sufficient for the purposes here.

The procedure\cite{shankar} is standard Wilsonian RG, where the field $\varphi$ is divided into a slow and a fast part,  $\varphi (\tau) = \varphi_s (\tau) + \varphi_f (\tau)$, with
\begin{eqnarray}
\varphi_s (\tau) &=& \frac{1}{2\pi} \int_{-\Lambda / b}^{\Lambda /b} \mathrm{d} \omega \ e^{-i\omega\tau} \varphi(\omega), \\
\varphi_f (\tau) &=& \frac{1}{2\pi} \int_{\Lambda / b < |\omega | < \Lambda } \mathrm{d} \omega \  e^{-i\omega\tau} \varphi(\omega) .
\end{eqnarray}
An effective action $S_s [\varphi_s]$ is then found for the slow field by integrating out the fast components,
\begin{equation}
e^{-S_s[\varphi_s]} = \int \mathcal{D} [ \varphi_f] \  e^{-S[\varphi]} .
\end{equation}
Rescaling the high-energy cutoff $\Lambda = v / \kappa$ with the parameter $b$, so that $\Lambda/b$ is the new cutoff, gives the new rescaled effective action and allows extraction of the RG flow equations.

\begin{figure*}[t!]
	\begin{center}
		\includegraphics[width=2.0\columnwidth]{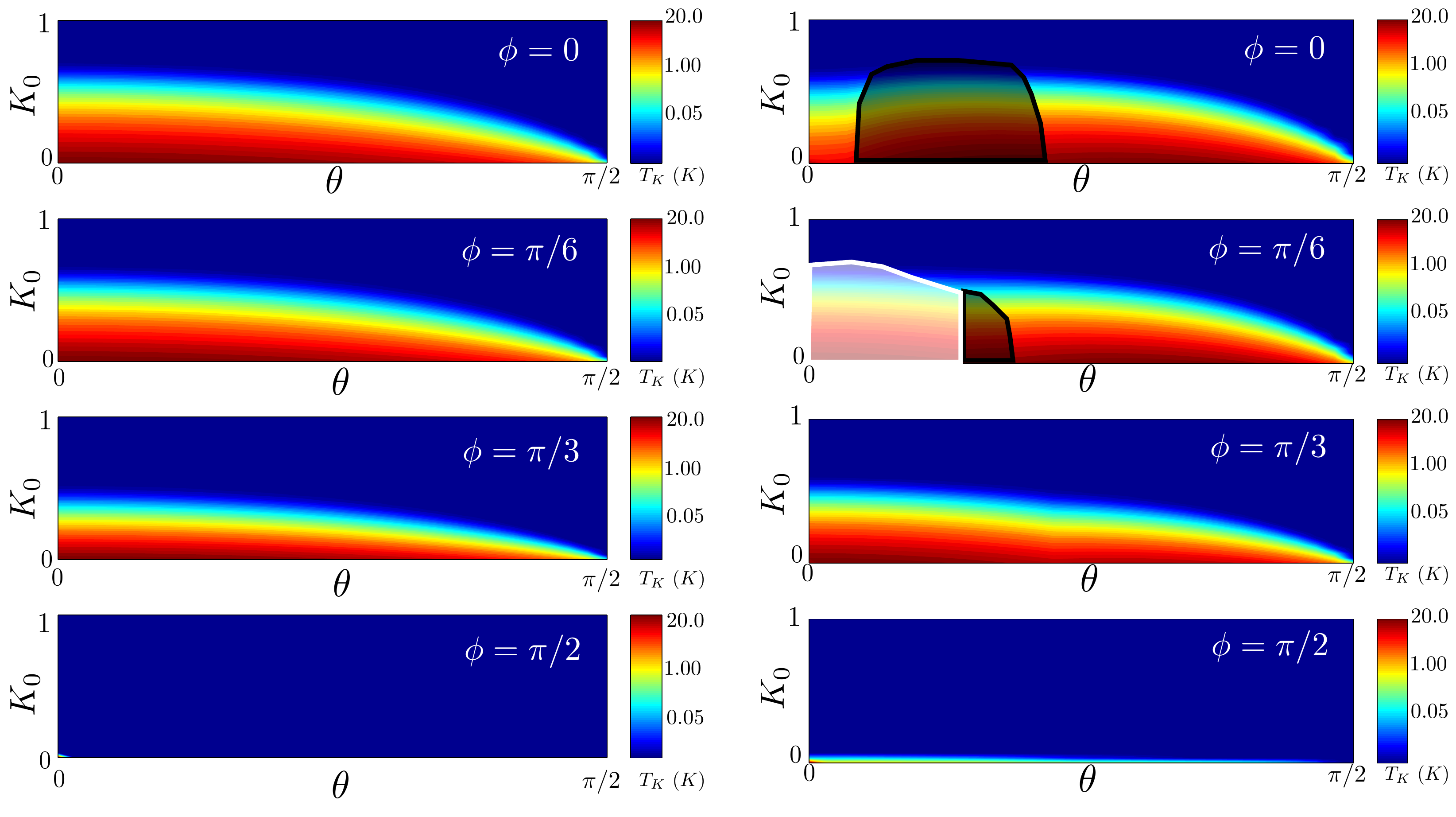}
		\caption{The Kondo temperature $T_K$ as a function of the Rashba angle $\theta$ and the ordinary Luttinger parameter $K_0$, for four different values of the Dresselhaus angle $\phi$. Note that the temperature scale is logarithmic, and that the vanishing Kondo temperature at $\theta,\phi =\pi /2$ simply reflect the diverging spin-orbit coupling strength in these limits. Left column:  $J_x = J_y \geq J_z$ (here with $J_x/a=J_z/a=10$ meV). Right column: $J_x  = J_y <J_z$ (here with $J_x/a=5$ meV and $J_z/a=50$ meV). In the black (white) shaded area, $J_{yz}$ ($J_{xz}$) dominates the perturbative RG flow, hence preventing the low-temperature formation of a Kondo singlet.}
		\label{KondoPlot}
	\end{center} 
\end{figure*}

A linked cluster expansion to second order in the Kondo couplings,
\begin{eqnarray}
 S_s[\varphi_s] &=& S_0 [\varphi_s] + \langle S_K [\varphi] \rangle_f \nonumber \\ 
&& - \frac{1}{2} \left( \langle S_K [\varphi] ^2 \rangle_f - \langle S_K [\varphi]  \rangle_f ^2 \right) ,
\end{eqnarray}
gives the rescaled effective action \cite{essj,anders}
\begin{eqnarray}
 S[\varphi] &=& S_0 [\varphi] +  \nonumber \\
&& \ \nonumber \\
&& + \int \mathrm{d} \tau \ \Big\{ \    b^{1-K}  \Big[  \frac{A}{\pi \kappa}\!\cos[\sqrt{4\pi K}\varphi (\tau)  ]   \nonumber \\ 
&& \ \nonumber \\
&& \qquad \qquad \qquad  + \frac{B}{\pi \kappa} \!\sin[\sqrt{4\pi K}\varphi (\tau) ] \, \Big]  \nonumber \\
&& \ \nonumber \\
&& +\left( b^2 - b^{2-2K} \right)    \   \frac{[A,B]}{2 v^2 \pi^{3/2} }  \partial_{\tau} \varphi (\tau)  \label{raction} \\ 
&& \ \nonumber \\
&& - \left( b - b^{1-K} \right)   \ i  \frac{[B,C]}{\kappa v  \pi^{2} } \cos[\sqrt{4\pi K}\varphi (\tau)  ]  \nonumber \\
&& \ \nonumber \\
&&  + \left( b - b^{1-K} \right)    \ i  \frac{[A,C]}{\kappa v  \pi^{2} } \sin[\sqrt{4\pi K}\varphi (\tau)  ] \, \Big\},  \nonumber 
\end{eqnarray} 
where only the most relevant operators are kept. Among the less relevant operators that are neglected is the correlated two-particle backscattering operator $ \cos[\sqrt{16\pi K}\varphi (\tau)  ]$. It is hence generated during the RG flow, which is expected in the presence of electron-electron interactions.\cite{crepin,gs} In order to determine the Kondo temperature of the strong-coupling crossover, only the Kondo scattering which is the most relevant process needs to be considered.

The commutators of the impurity operators (\ref{ABC}) are

\begin{eqnarray}
\left[A,B\right] &=& 2i \Big[ (J_{xy} J'_z - J_{xz} J_{zy} ) S^x \nonumber \\
&& \qquad   + (J_{xz} J_{zx} - J'_x J'_z ) S^y \label{AB} \\
&& \qquad   + (J'_x J_{zy} - J_{xy} J_{zx} ) S^z \Big],  \nonumber 
\end{eqnarray}

\begin{eqnarray}
\left[ B,C \right] &=& 2i \Big[ (J'_y J'_z - J_{yz} J_{zy} ) S^x \nonumber \\
&& \qquad   + (J_{yz} J_{zx} - J_{yx} J'_z ) S^y \label{BC}  \\
&& \qquad   + (J_{yx} J_{zy} - J'_y J_{zx} ) S^z \Big], \nonumber 
\end{eqnarray}

\begin{eqnarray}
\left[ A,C \right] &=& 2i \Big[ (J_{xy} J'_z - J_{xz} J_{zy} ) S^x \nonumber \\
&& \qquad   + (J_{xz} J_{zx} - J'_x J'_z ) S^y  \label{AC}\\
&& \qquad   + (J'_x J_{zy} - J_{xy} J_{zx} ) S^z \Big]. \nonumber 
\end{eqnarray}

\begin{figure}[ht!]
	\begin{center}
		\includegraphics[width=1.0\columnwidth]{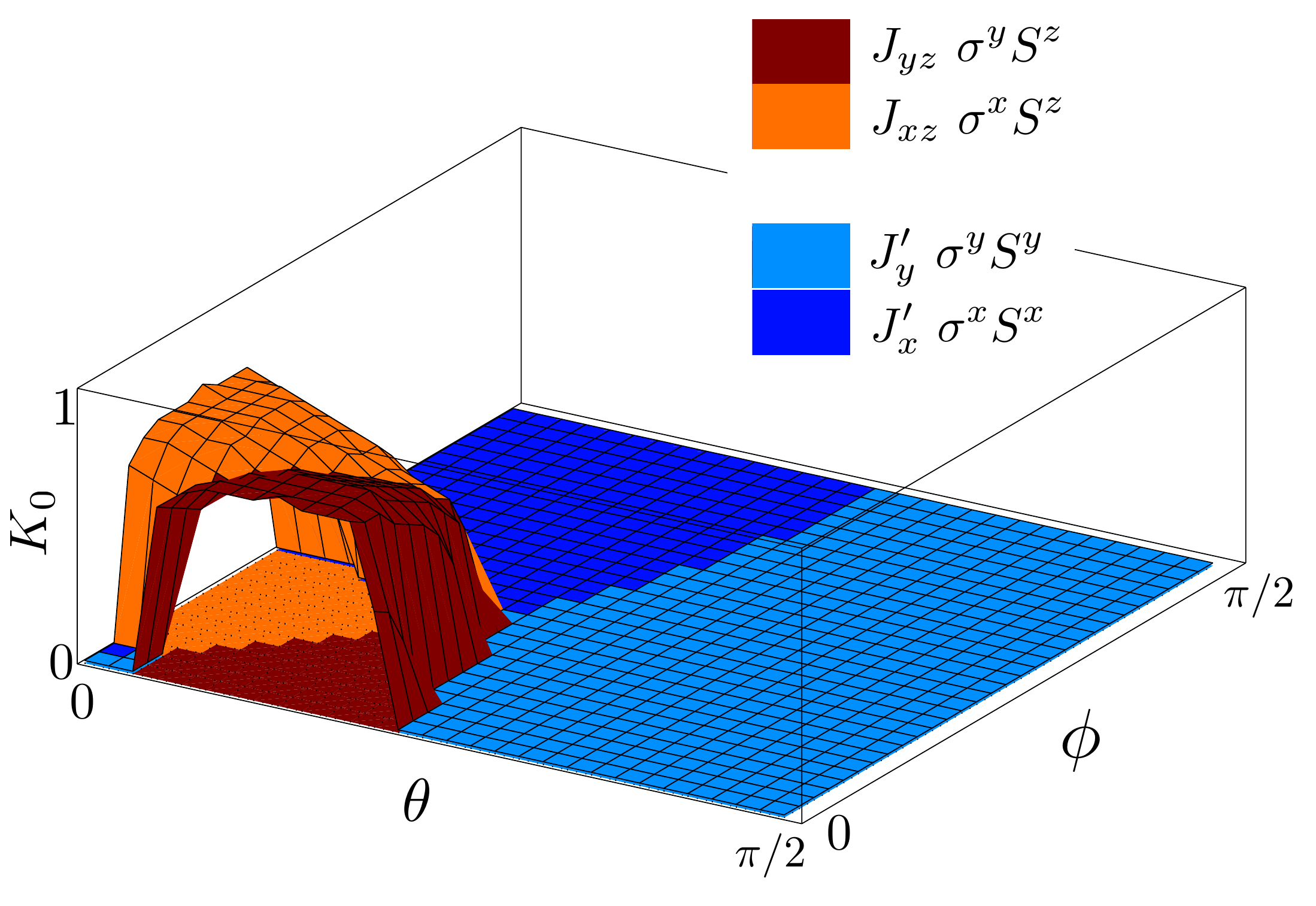}
		\caption{The "phase diagram", showing which Kondo coupling constant that dominates the low-temperature behavior when $J_x  = J_y <J_z$ [here with the same parameter values as for the right column in Fig.~(\ref{KondoPlot})]. Dark and light blue ($J'_{x}$ and $J'_y$, respectively) means the formation of a singlet between the electrons and impurity below the Kondo temperature. Red and orange ($J_{yz}$ and $J_{xz}$, respectively) denotes the "dome" within which the singlet formation is obstructed since instead the $J_{yz}\sigma^y S'^{z}$ or $J_{xz}\sigma^x S'^{z}$ interactions will dominate.}
		\label{3dPlot}
	\end{center} 
\end{figure}

Collecting the terms in Eq.~(\ref{raction}), using Eqs.~(\ref{AB})-(\ref{AC}), the RG equations for the Kondo couplings become
\begin{eqnarray}
\partial_{\ell} J'_x &=& (1-K ) J'_x + \nu K \left( J'_y J'_z - J_{yz} J_{zy} \right), \nonumber \\
\partial_{\ell} J'_y &=& (1-K ) J'_y + \nu K \left( J'_x J'_z - J_{xz} J_{zx} \right), \nonumber\\
\partial_{\ell} J'_z &=& \nu K \left( J'_x J'_y - J_{xy} J_{yx} \right), \label{rg} \\
\partial_{\ell} J_{xy} &=& (1-K ) J_{xy} + \nu K \left( J_{yz} J_{zx} - J'_z J_{yx} \right), \nonumber\\
\partial_{\ell} J_{yx} &=& (1-K ) J_{yx} + \nu K \left( J_{xz} J_{zy} - J'_z J_{xy} \right), \nonumber\\
\partial_{\ell} J_{xz} &=& (1-K ) J_{xz} + \nu K \left( J_{yx} J_{zy} - J'_y J_{zx} \right), \nonumber\\
\partial_{\ell} J_{zx} &=& \nu K \left( J_{xy} J_{yz} - J'_y J_{xz} \right),\nonumber \\
\partial_{\ell} J_{yz} &=& (1-K ) J_{yz} + \nu K \left( J_{xy} J_{zx} - J'_x J_{zy} \right), \nonumber\\
\partial_{\ell} J_{zy} &=&   \nu K \left( J_{xz} J_{yx} - J'_x J_{yz} \right), \nonumber
\end{eqnarray}
in terms of the "renormalization length" $\ell = \ln b$, and where $\nu = 1 / (\pi  v)$ is the density of states. Note that although the RG flows are different, the bare values of the coupling constants, given by Eq.~(\ref{HKp}) where $J_{xy}(\ell = 0) = J_{yx} (\ell = 0) $ etc., are not all independent.

The RG equations (\ref{rg}) are now solved numerically. As the temperature $T = D /b = D \, e^{-\ell} $ is reduced, the Kondo temperature $T_K$ is reached when one of the Kondo couplings in Eq.~(\ref{HKp}) becomes of the same order of magnitude as the energy scale of the unperturbed Hamiltonian (\ref{H0p}), i.e. $v_{\alpha \beta}$. Fig.~\ref{KondoPlot} shows $T_K$ as a function of the three independent parameters, the Rashba angle $\theta$, the Dresselhaus angle $\phi$, and the the ordinary Luttinger parameter $K_0 = K(\theta=0,\phi=0)$,
\begin{equation}
K_0 =  \sqrt{ \frac{\pi v_F }{ \pi v_F \!+2g} }
\end{equation}

\noindent where $g=g_f = g_d$ refers to the original basis. The plots show a Kondo temperature that for given $K_0$ and Dresselhaus angle $\phi$ depends on the Rashba angle, which is controllable via an electric field. For "easy-plane" and isotropic Kondo coupling ($J_x,J_y > J_z$ and $J_x=J_y=J_z$, respectively, which lead to very similar plots) the system flows towards the Kondo fixed point with divergent $J'_x$ or $J'_y$ couplings resulting in a singlet between the impurity and electrons. However, for "easy-axis" Kondo coupling ($J_x,J_y < J_z$), there are parameter regions where instead the couplings $J_{yz}\sigma^y S'^{z}$ or $J_{xz}\sigma^x S'^{z}$ dominate the RG flow, preventing singlet formation. These regions are indicated in Fig.~\ref{KondoPlot}, but note that their precise sizes depend on the numerical values of $J_x$,$J_y$ and $J_z$. In Fig.~(\ref{3dPlot}) they are shown for all values of $\phi$, $\theta$ and $K_0$, indicating that the "dome" is not completely symmetric under rotation in the $xy$-plane.

\section{Linear conductance}

\subsection{Strong-coupling limit at zero temperature}

The low-temperature limit of the conductance is governed by the RG flows of the Kondo (\ref{KondoBoson}), correlated two-particle (\ref{BackBoson}) and inelastic single particle (\ref{IEBoson}) backscattering terms in the Hamiltonian. The inelastic single-particle backscattering term $H'_{ie}$ has scaling dimension $2+K$, and is hence irrelevant in the sense of RG, contributing a term $\delta G_{ie} \sim T^{2+2K}$ to the conductance.\cite{Schmidt,Lezmy} The correlated two-particle backscattering term $H'_{2p}$ has scaling dimension $4K$ and is an irrelevant perturbation when $K > 1/4$, in which case it contributes a term $\delta G_{2p} \sim T^{8K-2}$ to the conductance.\cite{M} For $K<1/4$, the term $H'_{2p}$ turns relevant, hence opening up a gap at the energy scale of $T_{2p}$, leading to zero conductance of the system at $T=0$ which is approached as\cite{M} $G \sim T^{2(1/4K-1)}$.

Since the Kondo backscattering is a relevant (marginally relevant) perturbation for $K<1$ ($K=1$), the transport properties of the system in the zero-temperature limit when $1/4 \leq K \leq 1$ depend on what type of Kondo fixed point is reached. Outside the domes of dominating $J_{yz}\sigma^y S'^{z}$ or $J_{xz}\sigma^x S'^{z}$ interactions, the impurity is locked into a Kondo singlet below the energy scale of the Kondo temperature $T_K$. For a helical edge liquid of a two-dimensional topological insulator, this means that the Kondo impurity effectively disappears from the problem, with no contribution to the zero-temperature conductance,\cite{M} whereas a quantum wire would effectively be cut into two halves resulting in an insulating state. Inside the dome, where Kondo screening is obstructed resulting in a non-vanishing effective spin, the diverging scattering strength suggests the destruction of the gapless state below $T_K$.

\subsection{Linear response at weak coupling}

\subsubsection{The ac conductance}

For temperatures $T > T_K , T_{2p}$, the system is in a weak-coupling regime where the Kondo couplings $J$ and the two-particle backscattering coupling $g_{2p}$ remain small parameters. This allows a perturbative calculation of the linear conductance.
Since the components $ \psi'_{+}$ and $ \psi'_{-}$ of the spinor $\Psi'$ in the rotated basis are left- and right movers, the current operator $\hat{I}$ is given by 
\begin{equation} \label{current}
\hat{I}= (e/2)\,\partial_t \,( \Psi'^{\dagger} \sigma^{z} \Psi'). 
\end{equation}
Hence the part $\delta \hat{I}$ of the current operator that is due to the Kondo scattering $H_K^{\prime}$ is given by $\delta \hat{I} = e/(i 2 \hbar ) [  \Psi'^{\dagger} \sigma^{z} \Psi' , H'_K] $. It is here possible to conveniently consider the effects of the $J'_z$ coupling constant in Eq.~(\ref{HKp}) to all orders in perturbation theory by performing the unitary transformation $U (H'_{HLL}\!+\!H'_K) U^{\dagger}$, with \cite{ek}
\begin{equation}
U =  e^{i \lambda\varphi(0)S'^z},
 \end{equation}
  which cancels the $J'_z$ term in $H'_K$ against one that is generated in $H'_{HLL}$ provided one chooses $\lambda = J'_z/\pi v \sqrt{K}$. This transforms the Kondo Hamiltonian $H'_K $ to
\begin{eqnarray} 
H'_K  &=&  \frac{1}{2\pi\kappa}  \Bigg[  \left( \frac{J'_x + J'_y}{2} \right) e^{i \sqrt{\pi}(2\sqrt{K} -\lambda) \varphi} S'^+  \nonumber \\
&& +  \left( \frac{J'_x - J'_y}{2} + i \frac{J_{xy}+J_{yx}}{2} \right) e^{i \sqrt{\pi}(2\sqrt{K} +\lambda) \varphi} S'^- \nonumber  \\
&& +  \left( \frac{i J_{yz} + J_{xz}}{2} \right)  \, e^{ i \sqrt{4\pi K} \varphi} S'^z \Bigg] \label{UHU} \\
&&  \hspace{-0.4cm} +  \left( \frac{i J_{zy} + J_{zx}}{2} \right) \frac{1}{\sqrt{\pi K}}  : \partial_x\vartheta e^{ -i \sqrt{\pi }\lambda \varphi} : S'^{+}      \           + \mbox{H.c.}  \nonumber 
\end{eqnarray} 
in terms of the spin raising and lowering operators $S'^{\pm} = S'^x \pm i S'^y$. The current correction operator $\delta \hat{I} $ then takes the form
\begin{eqnarray}
\delta \hat{I}  &=&  \frac{i e}{2\pi\kappa}  \Bigg[  \left( \frac{J'_x + J'_y}{2} \right) e^{i \sqrt{\pi}(2\sqrt{K} -\lambda) \varphi} S'^+  \nonumber \\
&& +  \left( \frac{J'_x - J'_y}{2} + i \frac{J_{xy}+J_{yx}}{2} \right) e^{i \sqrt{\pi}(2\sqrt{K} +\lambda) \varphi} S'^-  \nonumber \\
&& +  \left( \frac{i J_{yz} + J_{xz}}{2} \right)  \, e^{ i \sqrt{4\pi K} \varphi} S'^z \ \Bigg]  + \mbox{H.c.}   \label{dI}
\end{eqnarray} 
Note that both $H'_K $ and $\delta \hat{I} $ are local at $x=0$, and for brevity only the time arguments of the fields are specified.
The Kubo formula for the conductance, $ G(\omega) = (1/ \hbar \omega) \int_{0}^{\infty} \mathrm{d} t \, e^{i\omega t} \langle \big[  I^{\dagger} (t) , I (0) \big] \rangle$, now gives the linear conductance correction $\delta G = G - G_0$ due to the Kondo scattering.
To second order in the Kondo couplings (except for $J'_z$ which is treated exactly), the unperturbed retarded Green's functions\cite{G}
\begin{eqnarray}
\langle \big[e^{i 2\sqrt{K_j} \varphi(t)} , e^{-i 2\sqrt{K_j} \varphi(0)} \big] \rangle_0  & =& \left( \frac{\pi \kappa}{\beta v}  \right)^{2K_j} 2 i \,\sin (\pi K_j)\nonumber \\
&& \times   \Big( \sinh \frac{\pi t }{\beta} \Big)^{-2K_j}
\end{eqnarray}
contribute three terms in $\delta G$ 
\begin{eqnarray}
\hspace{-0.5cm} \delta G(\omega) &=& - \sum_{j=1}^{3}  \frac{2e^2}{\hbar^3}  \frac{|A_j|^2}{(2\pi \kappa)^2}   \left( \frac{\pi T}{D}  \right)^{2K_j}\nonumber \\
&& \times   \sin (\pi K_j)  \frac{1}{i \omega} \int_0^{\infty}  \mathrm{d}t \ \frac{e^{-\mathrm{i} \omega t}-1}{(\sinh \frac{\pi  T t }{\hbar} )^{2K_j}}.
\end{eqnarray}
where $A_1 = (J'_x + J'_y)/2$, $A_2 = (J'_x- J'_y)/2 + i (J_{xy}+ J_{yx})/2$, $A_3 = (i J_{yz} + J_{xz})/2$, $K_{1,2}= K_{\mp} \equiv (\sqrt{K} \mp \lambda/2 )^2$ and $K_3 = K$. As in Refs.~\onlinecite{M} and \onlinecite{T}, the standard integrals are calculated by changing integration variable to $x = \pi T t / \hbar $ and then to $ u =  \textrm{tanh}\, x$, giving in the limit $\omega \ll T$
\begin{align} 
 &\frac{1}{i \omega}  \int_0^{\infty}  \mathrm{d}t \ \frac{e^{-\mathrm{i} \omega t}-1}{(\sinh \frac{\pi  T t }{\hbar} )^{2K_j}} \nonumber \\
& = \frac{\hbar}{\pi T} \frac{1}{\mathrm{i} \omega}  \int_0^{\infty}  \mathrm{d}x \ \frac{e^{-\mathrm{i} (\hbar \omega / \pi T) x}-1}{(\sinh x)^{2K_j}}\nonumber \\
 &\approx  \left(\frac{\hbar}{\pi T}\right)^2   \int_0^{\infty}  \mathrm{d}x \ \frac{x}{(\sinh x)^{2K_j}} \nonumber \\
  &= \left(\frac{\hbar}{\pi T}\right)^2  \int_0^{1}  \mathrm{d}u\  u^{-2K_j} (1-u^2)^{K_j-1} \textrm{arctanh}\, u,
  \end{align}
which is evaluated to give
\begin{equation} \label{dgt}
\delta G(\omega) =   - \frac{e^2}{ 2T}\,  \sum_{j=1}^{3}  |A_j|^2 \frac{1}{(\pi \hbar)^2 v \kappa} \frac{[\Gamma(K_j)]^2}{\Gamma(2K_j)}  \left( \frac{2\pi T}{D}  \right)^{2K_j-1} .
  \end{equation}
Hence $\delta G (\omega)$ is $\omega$-independent in the limit $J^2 \ll \omega \ll T$, where $J$ denotes the set of Kondo couplings. Rewriting Eq.~(\ref{dgt}) gives the final expression for the conductance correction from the Kondo scattering when $J^2 \ll \omega \ll T$ 
\begin{align}
&\hspace{-0.25cm} \delta G = -\frac{e^2}{\hbar} \frac{1}{4\pi (\hbar v)^2} \times \nonumber \\
 & \Bigg[ \  \left( \frac{J'_x + J'_y}{2} \right)^2 \frac{[\Gamma(K_-)]^2 }{\Gamma(2K_-)}  \left( \frac{2\pi T}{D}\right)^{2K_- - 2} \nonumber \\
& + \left|  \frac{J'_x - J'_y}{2} + i  \frac{J_{xy}+J_{yx}}{2}  \right|^2 \frac{[\Gamma(K_+)]^2 }{\Gamma(2K_+)}  \left( \frac{2\pi T}{D}\right)^{2K_+ - 2} \nonumber \\
& + \left|   \frac{i J_{yz} + J_{xz}}{2} \right|^2 \frac{[\Gamma(K)]^2 }{\Gamma(2K)}  \left( \frac{2\pi T}{D}\right)^{2K - 2} \ \Bigg]. \label{dG}
\end{align}

\subsubsection{The dc limit}

It is important to keep in mind that Eq.~(\ref{dG}) for the conductance correction was derived perturbatively in the Kondo couplings $J$. Hence when $\omega \ll  J^2 $ the above type of analysis is no longer valid, as noted in Ref.~\onlinecite{T} where instead a rate equation was set up for the impurity spin. In Ref.~\onlinecite{essj} this rate-equation approach was extended to incorporate the additional scattering types present when a Rashba interaction is added. Since a Dresselhaus interaction merely changes the different rates for the scattering events in these rate equations, the calculations in the supplemental material of Ref.~\onlinecite{essj} are easily modified to include also Dresselhaus interaction. For completeness, let us nevertheless carry them out explicitly also here.

Denoting the rates for the scattering events encoded in the Hamiltonian (\ref{HKp}) with $\gamma_{\pm}$ for $\sigma^{\mp} S'^{\pm}$, $\gamma'_{\mp}$ for $ \sigma^{\pm} S'^{\pm}$, $\gamma^z_{\pm}$ for $\sigma^{\mp} S'^{z} $ and $\tilde{\gamma}^z_0$ for $ \sigma^{z} S'^{\pm} $, the rate equation for the impurity spin becomes
\begin{equation} \label{re}
\partial_t P_{+} = (\gamma_+ + \gamma'_- + \tilde{\gamma}^z_0) P_{-} - (\gamma_- + \gamma'_+ + \tilde{\gamma}^z_0) P_{+}, 
\end{equation}
with $P_{\pm}$ the probability of the impurity spin being in the $+$ or $-$ state, and the probabilities obeying $P_+ + P_- = 1$. When adding a voltage bias $V = V_0 e^{-i \omega t}$ between the left- and right-moving electrons, encoded by the term
\begin{equation} \label{HV}
H_V = \frac{eV}{2} \int \mathrm{d}x \,\,\Psi'^{\dagger}  \sigma^{z}  \Psi'  
\end{equation}
in the Hamiltonian, the voltage-dependent rates are given by $\gamma_{\pm} = \gamma_0 \Lambda_{\pm}, \gamma^{\prime}_{\pm} = \gamma^{\prime}_0 \Lambda_{\pm}, \gamma^z_{\pm} = \gamma^z_0 \Lambda_{\pm}$,
with $\Lambda_{\pm} \equiv 1\pm eV/2T$ for $eV\ll T$. The current correction from the impurity scattering can now be written as 
\begin{eqnarray}
 \delta I &=&  -e \,\big( \gamma_+ P_{-} -\gamma'_- P_{-} + \gamma'_+  P_{+}  
- \gamma_- P_{+} \nonumber \\
&& \qquad \qquad + \gamma^z_+ /2 - \gamma^z_- /2  \big)  . \label{ri}
\end{eqnarray} 
Substituting the solution of Eq.~(\ref{re}) into Eq.~(\ref{ri}), one finds the linear conductance correction $\delta G(\omega) = \delta I / (V_0 e^{-\mathrm{i}\omega t})$ to be
\begin{eqnarray}
\delta G(\omega )  &=& -  \frac{e^2}{2T}    \big[ (\gamma_0+\gamma'_0 + \gamma_0^z)\omega + i8\gamma_0 \gamma'_0 \nonumber \\
&& \qquad \  + i2(\gamma_0+\gamma'_0)(\gamma_0^z +\tilde{\gamma}_0^z ) + i2\tilde{\gamma}_0\gamma_0^z  \big] \nonumber \\
&& \qquad    \times   \big[  i2(\gamma_0+\gamma'_0 + \tilde{\gamma}_0^z) + \omega \big]^{-1}  . \label{dGr}
\end{eqnarray}
In the dc limit, i.e. when $\omega \ll J^2 \ll T$, this becomes
\begin{align} 
& \hspace{-0.4cm} \delta G (\omega \to 0)=   \nonumber \\
&- \frac{e^2}{2T} \Big[ \frac{ 4\gamma_0 \gamma'_0\! +\!(\gamma_0\!+\!\gamma'_0)(\gamma_0^z \!+\!\tilde{\gamma}_0^z ) \!+\! 
\tilde{\gamma}^z_0\gamma_0^z}{\gamma_0\!+\!\gamma'_0\! +\! \tilde{\gamma}_0^z } \Big] .  \label{dGdc}
\end{align}
The remaining task is to determine the rates $\gamma_0$, $\gamma'_0$, $\gamma_0^z$ and $\tilde{\gamma}_0^z$. In the high-frequency regime one can compare Eq.~(\ref{dGr}), which gives $\delta G(\omega \gg \gamma)  = -(e^2 / 2T) (\gamma_0+\gamma'_0 + \gamma_0^z  )$, with the linear-response result in Eq.~(\ref{dG}). This gives
\begin{eqnarray}
\gamma_0 &\sim& (J'_x+J'_y)^2 (T/D)^{2(\sqrt{K}-\lambda/2)^2-1} \nonumber\\
\gamma_0^{\prime} &\sim& |J'_x- J'_y + i( J_{xy}+J_{yx})|^2 (T/D)^{2(\sqrt{K}+\lambda/2)^2-1} \nonumber \\
\gamma_0^z &\sim& | i J_{yz} + J_{xz}|^2 (T/D)^{2K-1}
\end{eqnarray}
In order to obtain the rate $\tilde{\gamma}_0^z$ of impurity spin-flips not accompanied by electron backscattering, one needs to do an analogous linear-response analysis for a field that instead couples to the impurity.\cite{essj} Such a calculation yields
\begin{eqnarray}
\tilde{\gamma}_0^z &\sim& | i J_{zy} + J_{zx} |^2 \,T/D
\end{eqnarray}
coming from the last term in Eq.~(\ref{UHU}).

\section{Backscattered current}

\subsection{I-V characteristics}

A finite time-independent bias voltage $V = V_0$ is conveniently treated in a picture where the voltage term (\ref{HV}) in the Hamiltonian is replaced by a time dependence in the electron fields, $\psi'_{\pm} \to e^{-i H_V t / \hbar} \psi'_{\pm}$. In this picture
\begin{eqnarray} 
H'_K(t)  &=&  \frac{1}{2\pi\kappa}  \Bigg[  A_1e^{i eV t/ \hbar}  e^{i \sqrt{\pi}(2\sqrt{K} -\lambda) \varphi} S'^+  \nonumber \\
&& +  A_2\, e^{i eV t/ \hbar}  e^{i \sqrt{\pi}(2\sqrt{K} +\lambda) \varphi} S'^- \nonumber  \\
&& +  A_3  \,e^{i eV t/ \hbar}   e^{ i \sqrt{4\pi K} \varphi} S'^z \label{UHUt} \\
&&  \hspace{-0.4cm} +  A_4 \frac{1}{\sqrt{\pi K}}  : \partial_x\vartheta e^{ -i \sqrt{\pi }\lambda \varphi} : S'^{+}      \         \Bigg]  + \mbox{H.c.}  \nonumber 
\end{eqnarray} 
and
\begin{eqnarray} \label{dIt}
\delta \hat{I}(t)  &=&  \frac{i e}{2\pi\kappa}  \Bigg[  A_1 e^{i eV t/ \hbar} e^{i \sqrt{\pi}(2\sqrt{K} -\lambda) \varphi} S'^+  \nonumber \\
&& +  A_2\, e^{i eV t/ \hbar} e^{i \sqrt{\pi}(2\sqrt{K} +\lambda) \varphi} S'^-  \\
&& + A_3 \, e^{i eV t/ \hbar} \, e^{ i \sqrt{4\pi K} \varphi} S'^z \ \Bigg]  + \mbox{H.c.}  \nonumber 
\end{eqnarray} 
where $A_1 = (J'_x + J'_y)/2$, $A_2 = (J'_x- J'_y)/2 + i (J_{xy}+ J_{yx})/2 $ and $A_3 = (i J_{yz} + J_{xz})/2$ as before, and $A_4 = (i J_{zy} + J_{zx})/2$.

\begin{figure*}[ht!]
	\begin{center}
		\includegraphics[width=1\columnwidth]{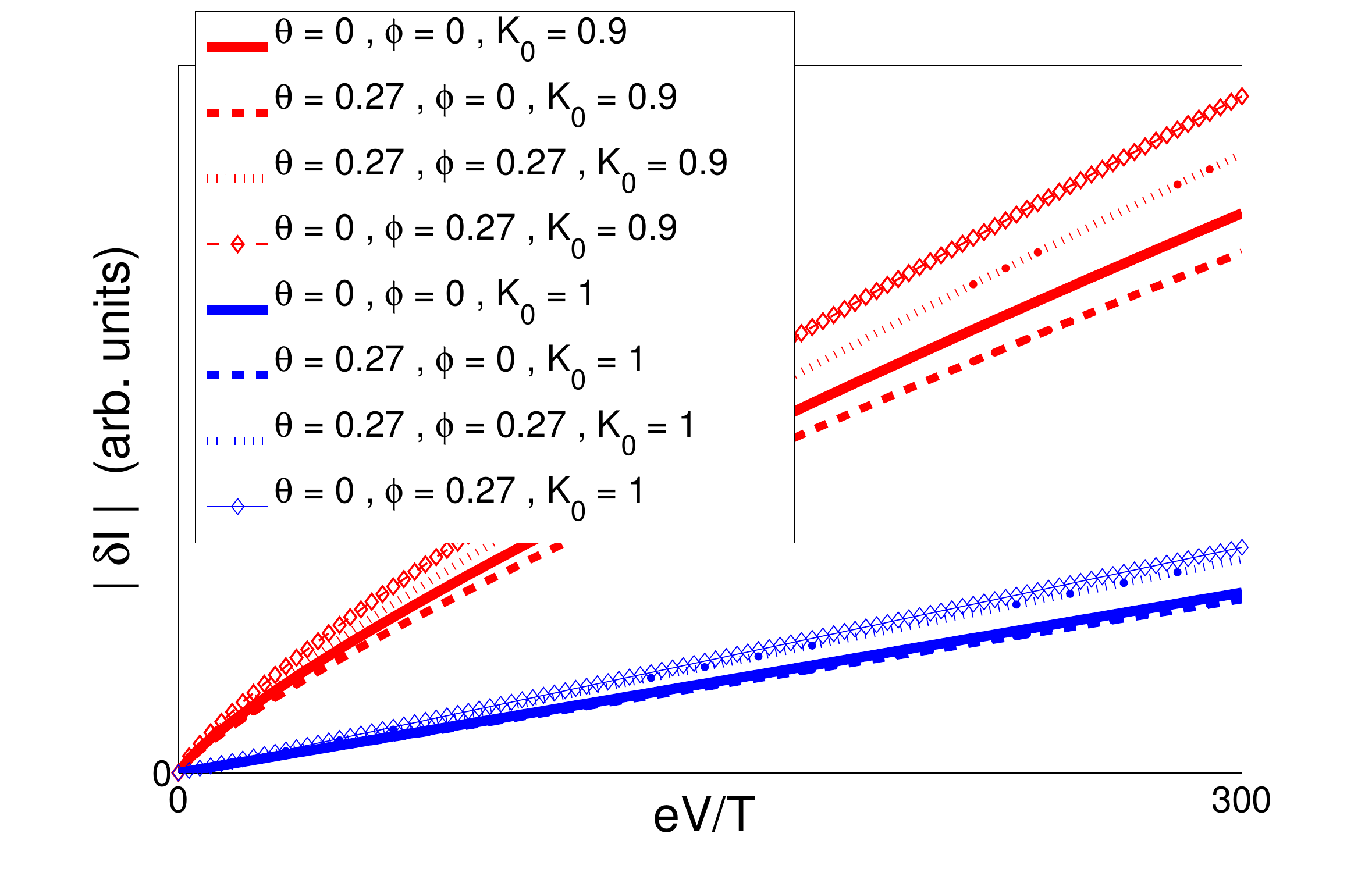}
		\includegraphics[width=1\columnwidth]{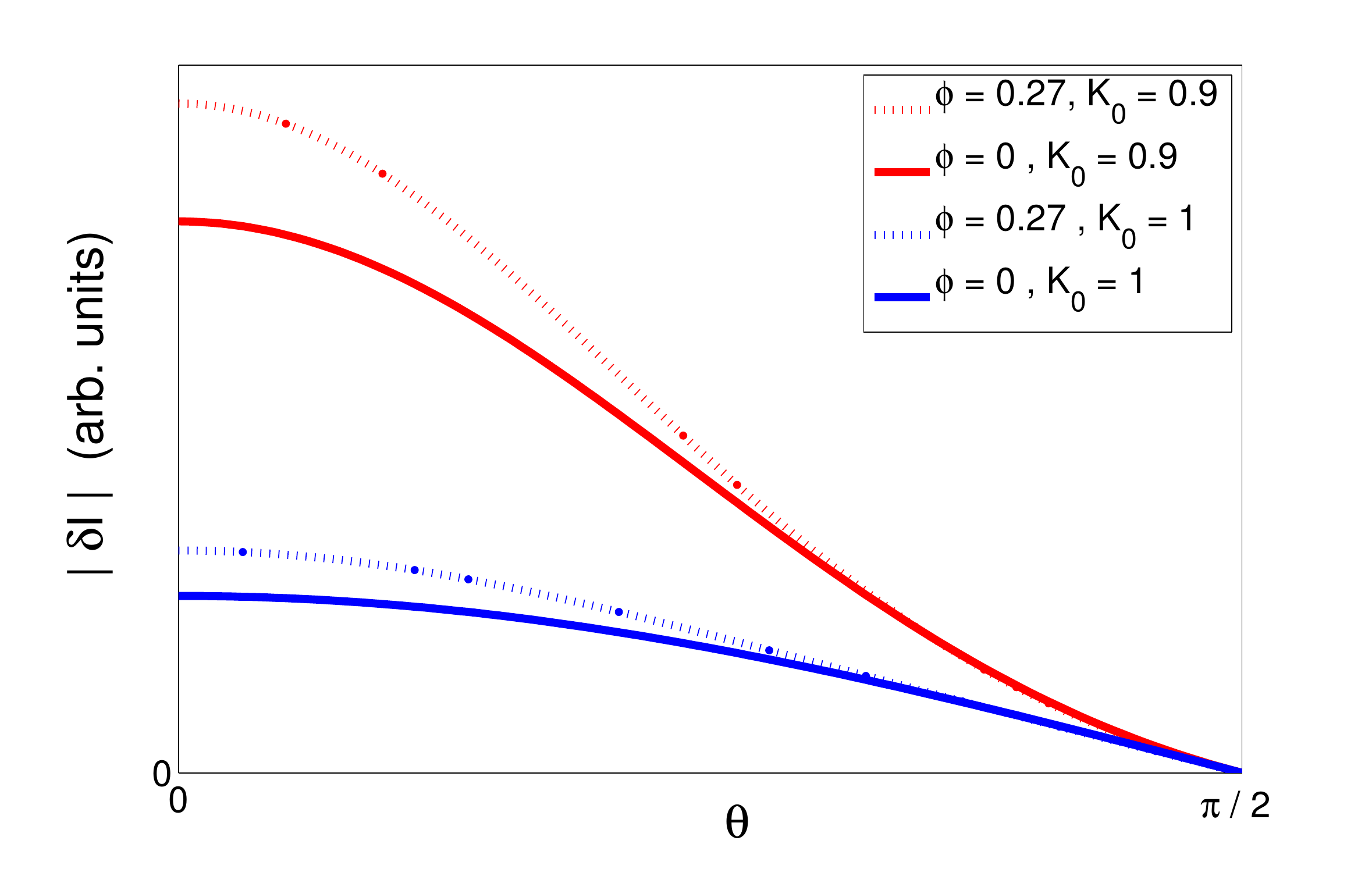}
		\caption{The backscattered current $\delta I$ due to the Kondo impurity, for different values of the Rashba and Dresselhaus angles $\theta$ and $\phi$ (parameterizing the strengths of the spin-orbit interactions), and the Luttinger parameter $K_0$. Left: Current-voltage characteristics for $\delta I$ at fixed temperature $T$. Right: $\delta I$ at fixed voltage and temperature, plotted as a function of the Rashba angle $\theta$. The coupling constants have been RG improved at the fixed temperature. With typical parameter values as discussed in the text, voltages on the order of 1 mV at temperatures in the mK regime will give current corrections $\delta I$ on the order of a nA.}
		\label{IVplot}
	\end{center} 
\end{figure*}

The non-equilibrium expectation value of the current operator is given by
\begin{equation}
\langle \delta \hat{I}(t) \rangle = \frac{1}{2} \sum_{\eta = \pm} \langle \hat{T}_K \left\{ \delta \hat{I}(t^{\eta}) \,e^{-i \int_K \mathrm{d} t' H'_K(t')} \right\} \rangle,
\end{equation}
where $\hat{T}_K$ is the time ordering operator on the Keldysh contour $K$, and $t^{\pm}$ is time $t$ either on the upper ($+$) or lower ($-$) branch of $K$, see e.g. Ref.~\onlinecite{martin}. To lowest order in the Kondo couplings, the most relevant terms are
\begin{eqnarray}
\langle \delta \hat{I}(t) \rangle  &=& - \frac{e}{8\pi^2 \kappa^2}  \sum_{\eta = \pm} \int_K \mathrm{d} t'  e^{i eV (t-t') / \hbar} \, \Bigg[     \langle  S'^+ S'^- \rangle   \nonumber \\ 
&& \hspace{-1.8cm} \times  \Big( \,  | A_1 |^2  \langle \hat{T}_K \{  e^{ i \sqrt{4\pi K_-} \varphi(t^{\eta}) }   e^{- i \sqrt{4\pi K_-} \varphi(t') } + H.c.  \} \rangle  \nonumber \\
&& \hspace{-1.8cm} +  | A_2 |^2  \langle \hat{T}_K \{  e^{ i \sqrt{4\pi K_+} \varphi(t^{\eta}) }   e^{- i \sqrt{4\pi K_+} \varphi(t') } + H.c.  \} \rangle \, \Big)  \label{dIt2} \\
&& \hspace{-1.8cm}+    | A_3 |^2  \langle \hat{T}_K \{  e^{ i \sqrt{4\pi K} \varphi(t^{\eta}) }   e^{- i \sqrt{4\pi K} \varphi(t') } + H.c.  \} \rangle  \langle  S'^z S'^z \rangle  \Bigg] \nonumber .
\end{eqnarray}

\noindent The integrals on the Keldysh contour are found\cite{martin} to result in
\begin{eqnarray}
\langle \delta \hat{I}(t) \rangle  &=& -\frac{e}{2 \pi^2 v \kappa}  \sinh \left( \frac{eV }{2T} \right) \sum_{j=1}^3 \Bigg[ | A_j|^2  \left( \frac{2\pi T}{D} \right)^{2K_j -1} \nonumber \\
&& \qquad \quad \times \frac{ \big| \Gamma\left( K_j + i eV / 2\pi T \right) \big|^2 }{\Gamma ( 2K_j)} \, \Bigg],  \label{dIV}
\end{eqnarray}
where $\Gamma$ is the gamma function. This could equivalently be written in terms of the Euler beta function $B$ as
\begin{eqnarray} \label{ivc}
\delta I\  &=&  \frac{e}{2 \pi^2 v \kappa} \sum_{j=1}^{3}  \mbox{Im}\Big\{ \, | A_j|^2  \left( \frac{2\pi T}{D} \right)^{2K_j -1} \nonumber \\
&& \times B(K_j + i e V/ 2\pi T , K_j -i e V / 2\pi T  )  \nonumber \\ 
&&  \times  \frac{ \sin[\pi(K_j -i e V / 2\pi T )] }{ \cos (\pi K_j)}  \Big\} \quad 
\end{eqnarray}
which is on the form of the result presented in Ref.~\onlinecite{essj}, derived using the tunneling formalism.\cite{Mahan,W} Note the difference in sign of the prefactors. Since the backscattering current $\delta I \equiv I - I_0 $, where $I_0$ is the current in the absence of the impurity, it is negative.
The current corrections due to correlated two-particle, and inelastic single-particle, backscattering can be found in Ref.~\onlinecite{Lezmy}.

In Fig.~\ref{IVplot} the backscattered current $\delta I$, given by Eq.~(\ref{dIV}), is plotted for different strengths of the Rashba and Dresselhaus interactions and Luttinger parameters $K_0$. As an illustration, possible parameter values can be collected from Ref.~\onlinecite{essj} for the case of an Mn$^{2+}$ ion implanted in the helical edge liquid of a HgTe quantum well.\cite{Datta} For this material,\cite{Konig} $v_F \approx 5.0 \times 10^5$ m/s, the lattice constant\cite{Efros} $a \approx 0.5$ nm, and the bandwidth\cite{Konig} $D \approx 300$ meV. Estimates for the value of the Luttinger parameter $K_0$ for the electron-electron interaction strength vary between\cite{M,Strom,Hou,Strom2,TeoKane} 0.5 and 1, depending on the geometry and composition of the specific semiconductor heterostructure. The Kondo coupling constants are taken to be $J_x = J_y = 3J_z = 3 J_I$, where a rough estimate yields\cite{essj,Ujsaghy,Zitko,Furdyna} $ J_I/a $ on the order of 10 meV. For HgTe quantum wells, values of the Rashba coupling $\hbar \alpha$ are typically in the range between \cite{Hinz} $5\times 10^{-11}$ eVm to $1 \times 10^{-10}$ eVm, i.e. the value of the Rashba rotation angle is roughly in the interval $0.14 < \theta < 0.27$ for typical samples and voltages. For this particular heterostructure, the Dresselhaus interaction strength is typically negligible compared to the Rashba,\cite{silsbee} but note there are other materials with helical edge states where Dresselhaus is comparable in magnitude or even stronger than the Rashba interaction.\cite{cw,my} With these parameter values, a voltage bias in the mV regime at a temperature around 1 mK typically yields a current $\delta I$ on the order of 1 nA, i.e. a correction of roughly one percent to $I_0 \equiv G_0 V$.

\subsection{Current fluctuations}

From the above results it is an easy task to extract the noise in the current correction $\delta I$. In terms of the symmetric combination of the noise correlators
\begin{eqnarray}
S(t-t') &=& \langle \delta \hat{I} (t) \delta \hat{I} (t') \rangle + \langle \delta \hat{I} (t') \delta \hat{I} (t) \rangle \nonumber \\
&& \qquad - 2\langle \delta \hat{I} (t') \rangle \langle \delta \hat{I} (t) \rangle, 
\end{eqnarray}
which in the Keldysh formalism takes the form
\begin{eqnarray} \label{Stt}
S(t-t') &=&  \sum_{\eta = \pm} \langle \hat{T}_K \left\{ \delta \hat{I}(t^{\eta}) \delta \hat{I}(t'^{-\eta})  \,e^{-i \int_K \mathrm{d} t' H'_K(t')} \right\} \rangle\nonumber \\
&& \qquad \qquad - 2 \langle \delta \hat{I} (t) \rangle,
\end{eqnarray}
the noise spectrum $S(\omega)$ follows as the Fourier transform $S(\omega) = \int_{-\infty}^{\infty} \mathrm{d} s\, e^{i\omega s} \, S(s)$; for a review see Ref.~\onlinecite{martin}. To lowest order in the Kondo couplings, Eq.~(\ref{Stt}) reduces to a current-current correlator, $S(t-t') =   \sum_{\eta = \pm} \langle \hat{T}_K \{ \delta \hat{I}(t^{\eta}) \delta \hat{I}(t'^{-\eta}) \} \rangle - 2 \langle \delta \hat{I} (t) \rangle $, which is evaluated precisely as Eq.~(\ref{dIt2}). This yields the zero-frequency noise
\begin{eqnarray}
S(\omega = 0)  &=& \frac{e^2}{ \pi^2 v \kappa}  \cosh \left( \frac{eV }{2T} \right) \sum_{j=1}^3 \Bigg[ | A_j|^2  \left( \frac{2\pi T}{D} \right)^{2K_j -1} \nonumber \\
&& \qquad \quad \times \frac{ \big| \Gamma\left( K_j + i eV / 2\pi T \right) \big|^2 }{\Gamma ( 2K_j)} \, \Bigg],
\end{eqnarray}
i.e. $S(\omega = 0) = 2e | \langle \delta \hat{I}(t) \rangle  | \coth (eV/2T)$. Since $\coth (eV/2T) \approx 1$ for the temperature and voltages used in Fig.~\ref{IVplot}, this gives a Fano factor $S(\omega=0) / (2e | \langle \delta \hat{I} \rangle |) \approx 1$ for the current $\delta I$ due to $H'_K$. The noise in the current corrections due to correlated two-particle, and inelastic single-particle, backscattering can be found in Ref.~\onlinecite{Lezmy}. In total, the combined backscattering current due to $H'_K + H'_{2p} + H'_{ie}$ has an enhanced Fano factor (between 1 and 2), due to the larger effective charge $e^* = 2e$ appearing in the two-particle backscattering noise.~\cite{sook}

\section{Thermal transport}

Let us also consider the thermal conductance of the system for temperatures $T > T_K$. In terms of left-and right-moving fields $\phi_{L,R} = (\vartheta \pm \varphi)/2$, such that $H'_{HLL} = (\hbar v/2)\int \textrm{d}x [ (\partial_x\phi_R)^2 + (\partial_x \phi_L)^2 ] = H_{HLL}'^{R} + H_{HLL}'^{L}$, the backscattered thermal current $\delta \hat{I}_Q $ is given by the operator \cite{KF}
\begin{eqnarray}
\delta \hat{I}_Q &=& \int \mathrm{d} x \, \partial_t (H_{HLL}'^{R} - H_{HLL}'^{L})/2  \\ 
&& \quad = \frac{iv}{ 2}  \int \mathrm{d} x  \left[ :\!\partial_x \varphi\, \partial_x \vartheta\!:\!(x) \, , \, H'_K(0) \right], \nonumber
\end{eqnarray}
which gives 
\begin{eqnarray} \label{dIQ}
\delta \hat{I}_Q  &=& - \frac{\hbar v \sqrt{\pi K}}{e}  \partial_x \varphi\,  \delta\hat{I} \\
&& + A_4 \frac{\sqrt{\pi} v}{2 \sqrt{K}}   \left(  : \partial_x^2 \vartheta \,e^{ -i \sqrt{\pi }\lambda \varphi} : S'^{+}    + \mbox{H.c.}  \right), \nonumber
\end{eqnarray}
where $H'_K$ and $\delta\hat{I}$ are given by Eqs.~(\ref{UHU}) and (\ref{dI}), and $A_4 =  (i J_{zy} + J_{zx})/2$. The first term in Eq.~(\ref{dIQ}) has its analog in Refs.~\onlinecite{KF} and \onlinecite{T}, but the second term is new and is a result of the spin-orbit interaction. The correction $\delta \mathcal{K}$ to the (zero-frequency) linear thermal conductance due to the Kondo impurity is obtained from the Kubo formula, $ \delta \mathcal{K} = \lim_{\omega \to 0} (1/ \hbar \omega T) \int_{0}^{\infty} \mathrm{d} t \,e^{i \omega t} \langle \big[  \delta \hat{I}_Q ^{\dagger} (t) , \delta \hat{I}_Q  (0) \big] \rangle$, where the two terms in $\delta \hat{I}_Q $ add separately to $\delta \mathcal{K}$. Now, the contribution from the first term in Eq.~(\ref{dIQ}) follows from Ref.~\onlinecite{KF}. The contribution from the second term follows by noting the close resemblance between that operator, with scaling dimension $2+\lambda^2/4$, and the inelastic single-particle backscattering term in Eq.~(\ref{IEBoson}), with scaling dimension $2+K$. This makes it possible to perturbatively evaluate the contribution from the second term in Eq.~(\ref{dIQ}) using known results from Ref.~\onlinecite{Lezmy}, valid at frequencies $\omega \gg J^2$. It follows from those results that this operator contributes a term $\sim\!(k_B T/D)^{\lambda^2 /2 + 1}$ to $ \delta \mathcal{K}$. In comparison, this contribution is subleading as $k_B T \ll D$, and the thermal conductance correction is hence given by the contribution from the first term in Eq.~(\ref{dIQ}), which gives the usual formula,\cite{KF,T}
\begin{eqnarray} 
\delta \mathcal{K} &\approx& \frac{\hbar^3 K}{8 e^2 k_B T^2} \int \mathrm{d} \omega \, \frac{\omega^2 \, \mathrm{Re}\, \delta G(\omega)}{\sinh^2 (\hbar \omega / 2k_B T)},
\end{eqnarray}
for the zero-frequency thermal conductance correction. The same observation as in Ref.~\onlinecite{T} therefore applies, namely that $\delta \mathcal{K}$ is determined by the electrical conductance correction $\delta G (\omega)$ within a frequency range $0 \leq \omega \lesssim 2 k_B T/ \hbar$, and that even when $\delta G$ vanishes in the dc limit this is not necessarily true for $\delta \mathcal{K}$.

\section{Discussion}

The interplay between a magnetic impurity and spin-orbit interaction in a helical Luttinger liquid has been studied. Using a transformation\cite{VO} that effectively removes the spin-orbit interactions among the electrons, at the price of introducing additional coupling terms to the impurity, the results are non-perturbative in the spin-orbit couplings. The effect described in Ref.~\onlinecite{essj}, that a Rashba spin-orbit interaction appears to prevent the Kondo singlet formation in certain parameter regimes, is found to persist also with a Dresselhaus spin-orbit interaction. The perturbative RG analysis therefore suggests that a time-reversal invariant perturbation can destroy the Kondo effect in one dimension, in stark contrast to the case for ordinary metals where the Kondo effect is robust,\cite{Meir,Malecki} and such perturbations only shift the Kondo temperature.\cite{Schuricht,Feng,Zitko2,Zarea, Isaev}

In the other parameter regimes, the Kondo effect will set in. Since the Rashba interaction can be controlled by an external gate voltage, this offers a mechanism to electrically control the Kondo effect and therefore the transport properties of the edge at low temperatures.\cite{essj} At higher temperatures the Rashba dependence of the Kondo scattering permits a way to electrically control the small current corrections. For in-plane symmetric Kondo couplings $J_x = J_y \neq J_z$, and negligible Dresselhaus interaction, Eq.~(\ref{dGdc}) reveals a vanishing dc linear conductance correction which can be made non-zero by tuning the gate voltage. With non-zero Dresselhaus interaction strength, the linear conductance correction however remains non-zero also in the dc limit even without Rashba interaction.

Unfortunately, there is no exactly solvable Toulouse limit as in the absence of spin-orbit interactions.\cite{T} This rules out an exact analysis of the low-frequency behavior at some special value for the interaction strengths, and has its root in the new operators generated by the spin-orbit interaction bringing several different scaling dimensions into the problem. The different exponents show up in the temperature dependence of conductances, and in the temperature and voltage dependence of the current correction due to the Kondo impurity. 

The spin-orbit interactions hence introduce new types of scattering mechanisms compared to those for the original helical liquid with a Kondo impurity, adding yet another piece to the different types of correlations appearing in topological insulator materials,\cite{ha} including the effects on Kondo screening,\cite{Wu,M,M2,T} transport properties\cite{M,T,KF} and current noise correlations.\cite{ss,llc} There is also the possibility of exotic multi-channel Kondo physics, like the two-channel effect appearing with a quantum dot between two helical edges\cite{law,posske} where the effects of a Rashba coupling were studied very recently.\cite{ll} Interesting extensions include the case of many impurities\cite{lp} forming a Kondo lattice\cite{M2} on a helical edge, where the effects of spin-orbit interactions remain unexplored.

\section*{ACKNOWLEDGMENTS}
It is a pleasure to thank Henrik Johannesson and Anders Str\"om for very valuable discussions. This research was supported by the Swedish Research Council (Grant No. 621-2011-3942).

\end{document}